\documentclass{article}

\usepackage[final, nonatbib]{tccml_neurips_2020}

\usepackage[utf8]{inputenc} 
\usepackage[T1]{fontenc}    
\usepackage{hyperref}       
\usepackage{url}            
\usepackage{booktabs}       
\usepackage{amsfonts}       
\usepackage{nicefrac}       
\usepackage{microtype}      
\usepackage{amsmath}
\usepackage{amssymb}
\usepackage{wrapfig}
\usepackage{floatrow}
\usepackage{graphicx}
\usepackage{float}
\usepackage{xcolor}
\usepackage{enumitem}

\def\x{{\mathbf{x}}}

\usepackage[utf8]{inputenc} 
\usepackage[T1]{fontenc}    
\usepackage{hyperref}       
\usepackage{url}            
\usepackage{booktabs}       
\usepackage{amsfonts}       
\usepackage{nicefrac}       
\usepackage{microtype}      

\title{Wildfire Smoke and Air Quality: How Machine Learning Can Guide Forest Management}

\author{%
	Lorenzo~Tomaselli  $^{1}$ $\quad\quad$ Coty Jen$^{\,2}$ $\quad\quad$  Ann B. Lee$^{\, 1}$ \\
	\\
	$^{1}$Department of Statistics \& Data Science, Carnegie Mellon University \\
	$^{2}$Department of Chemical Engineering, Carnegie Mellon University \\
	\texttt{ltomasel@andrew.cmu.edu} 
}

\begin{document}

\maketitle

\begin{abstract}
  Prescribed burns are currently the most effective method of reducing the risk of widespread wildfires, but a largely missing component in forest management is knowing which fuels one can safely burn to minimize exposure to toxic smoke. Here we show how machine learning, 
  such as spectral clustering and manifold learning, can provide interpretable representations and powerful tools for differentiating between smoke types, hence providing forest managers with vital information 
  on effective strategies to reduce climate-induced wildfires
  while minimizing production of harmful smoke.
\end{abstract}

\section{Introduction} 

As evident from the past decade, wildfires in the western US are becoming larger, deadlier, and more frequent due to worsening climate conditions, causing phenomenons such as recurring droughts, heat waves, and earlier springs \cite{Dennison14, Jolly2015, Westerling940}.
In an effort to reduce wildfire risk, state and local governments are pushing to manage forests via regular prescribed burning, i.e., controlled, low-temperature burns to reduce the amount of fuel in a forest.
Although prescribed burning conducted every few years will reduce wildfire risk due to climate change, populations near heavily forested areas will experience frequent smoke episodes. Previous research has observed that smoke produced from different burn conditions or from burning different fuels, such as whole trees and man-made structures in wildfires versus ground cover duff in prescribed burns, is
chemically distinct \cite{acp-11-4039-2011, Jen19}.
Specifically, we have previously demonstrated that burning a specific shrub, manzanita, can release enormous amounts of a toxic compound known as hydroquinone into the air \cite{Jen18}. Other smoke compounds that describe a specific fuel or burn condition may exist but there is currently no technique that can extract this information from the complex chemical composition.  Consequently, 
current air quality models treat all smoke as composed of about ten common compounds (i.e., produced from burning any vegetation) and overlooks differences between thousands of other compounds \cite{Wiedinmyer11}.
The lack of more fuel-specific smoke information prevents forest managers from developing an effective strategy to reduce climate-induced wildfire risk while minimizing air quality damage. 

A first step toward understanding the impact of wildland burns, including wildfires and prescribed burns, on air quality is through {\em chemical fingerprinting} ---  laboratory analysis to quantify compounds in collected smoke samples from wildland burns. The output from advanced analytic platforms, such as two-dimensional gas chromatography (GC) and mass spectrometry (MS), is multi-dimensional and contains thousands of chemical compounds \cite{Goldstein08}.
Comparing smoke samples and relating fingerprint data to their impacts on air quality, and subsequent exposure toxicity, is also an inherently difficult task because of complex synergism between numerous compounds and the variability in burn conditions. There is an urgent need for machine learning techniques that can differentiate between different types of smoke with interpretable and explainable results. We do not want to limit ourselves to only answering whether two smoke samples are different, but we aim at gaining insight as to how differences (if observed and statistically significant) relate to groups of compounds. In this way we can link smoke to fuel, burn conditions, and air quality impacts.

In this work, we describe how spectral clustering and manifold learning can provide interpretable representations and powerful tools for differentiating between smoke. The statistical challenge is that the data (such as mass spectra representing chemical compounds) are not only high-dimensional, but that each smoke sample is a weighted set of high-dimensional data. Furthermore, not all mass spectra are equally likely to occur; the underlying physical constraints naturally imply that the spectra, if seen as points in a higher-dimensional space, inherently possess ``sparse'' low-dimensional structure (clusters, manifolds, etc). The main question is how to design efficient representations and machine learning (ML) algorithms for smoke samples that take sparse structure into account, and that allow us to easily relate ML prediction results back to groups of chemical compounds. Here we present a new geometry-based metric between smokes as distributions over chemical compounds, which can serve as input to kernel ML and visualization algorithms.

\section{Data}
Our research team collected a total of 54 prescribed burn smoke samples at ground level and above the forest canopy ($\sim$100 m in the air). Burns were conducted at Blodgett Forest Research Station (BFRS, Georgetown, CA) in 2017. Additional $\sim$100 smoke samples will be collected during prescribed burns at BFRS in Nov. 2020. These field samples represent unknown fuel composite samples as we only know the approximate fuel composition prior to the prescribed burn. 
We will also conduct laboratory burns of fuels harvested from BFRS in order to separate chemical fingerprints of smoke produced from single and multicomponent fuel mixtures. 

Smoke samples are analyzed using two-dimensional GC-MS with online derivatization \cite{Jen19}. This instrument separates compounds within smoke particles by first separating molecules by their boiling point and then by their polarity before detection via electron ionization with a time-of-flight mass spectrometer. A compound is described by its retention time for boiling point (i.e., the time needed to travel through the first GC column), retention time for polarity, and mass spectrum. Approximately 1000 unique compounds were separated and quantified in each smoke sample. 
We define the {\em chemical fingerprint} of a smoke sample as a weighted set of compounds $S=\{(\mathbf{x}_1, w_{\mathbf{x}_1}),\dots,(\mathbf{x}_m, w_{\mathbf{x}_m})\}$, where $\mathbf{x}_1, \dots, \mathbf{x}_m \in \mathbb{R}^p$ represent $m \sim 1000$ chemical compounds in the sample as measured by MS, GC, or both, with dimension $p \sim 500$; the weights $w_{\mathbf{x}_1},\dots, w_{\mathbf{x}_m}\in [0,1]$ with $\sum_{i=1}^m w_{\mathbf{x}_i}=1$ represent emission intensities or mass concentrations ($mg/m^3$).

\section{Interpretable Insights on Smoke via Geometry-Based Metrics}

\begin{figure}[b]
	\floatbox[{\capbeside\thisfloatsetup{capbesideposition={right,top},capbesidewidth=.3\textwidth}}]{figure}[\FBwidth]
	{\caption{\small Toy example to illustrate Aim 1. In (a), we define a codebook for 
			chemical compounds via spectral clustering. Panel (b), bottom left, illustrates that a distance metric that does not utilize data geometry will place histogram representations $f$, $g$, and $h$ of smoke samples at equal distance, whereas the arrangement to the bottom right (implied by GDD) 
			better reflects that $f$ is closer to $g$ than to $h$.
		}
		\label{fig_methods}}
	{\fbox{\includegraphics[clip, trim=0cm 1.2cm 0cm 2.25cm, width=.65\textwidth]{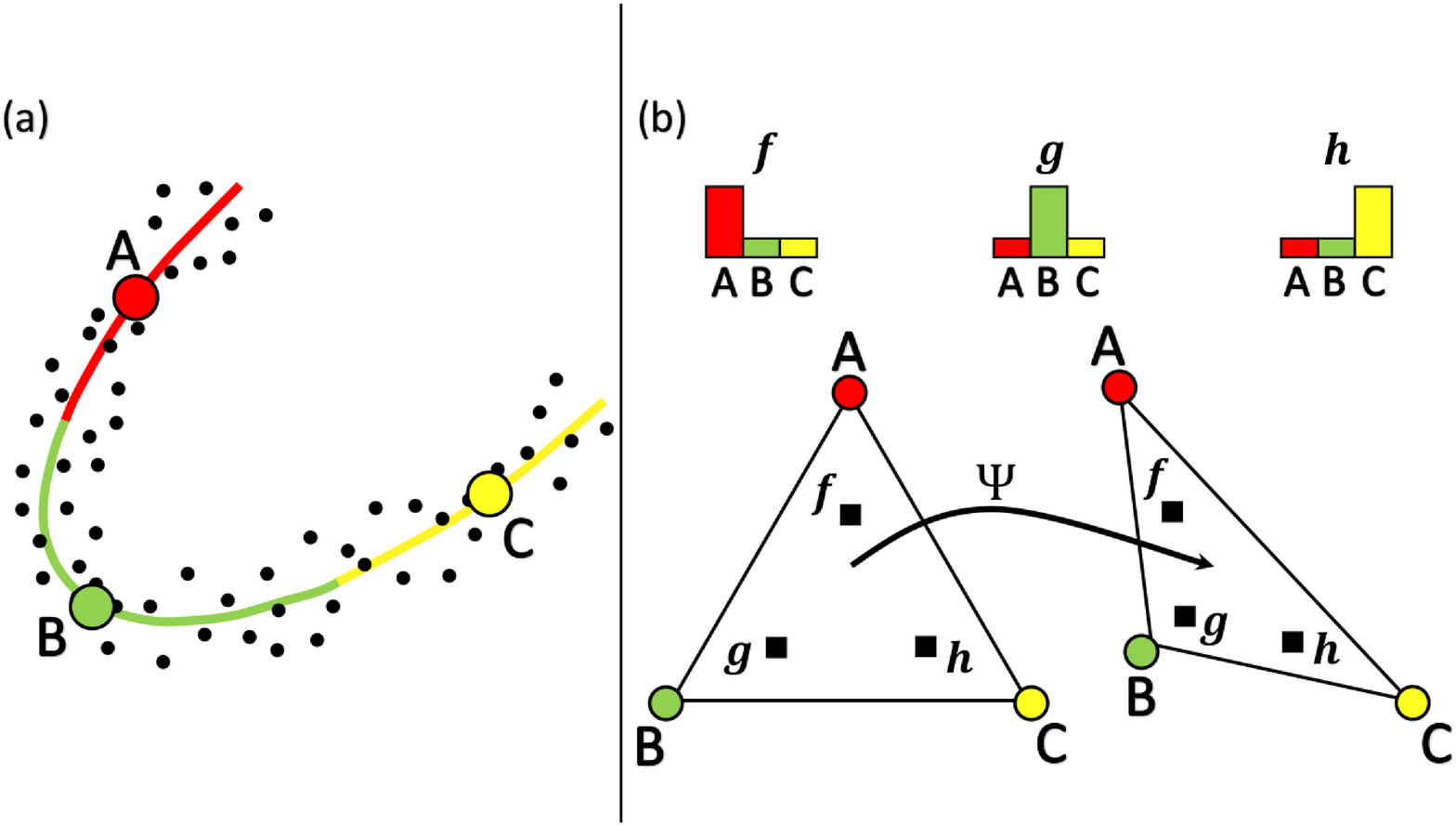}}}
\end{figure}

\subsection*{Aim 1: Building interpretable metrics that differentiate between smoke via manifold learning}

Although the data  $\x\in \mathbb{R}^p$ (for, e.g., mass spectra of chemical compounds)  are high-dimensional,
they have low intrinsic dimensionality due to physical constraints in the system. {\em How do we build interpretable metrics that reflect this structure with scalable ML algorithms for computing distances between smoke samples?} The toy example in Figure \ref{fig_methods} illustrates our proposed approach for comparing and analyzing sets of such high-dimensional data.  Panel (a) shows data lying near a low-dimensional manifold.  Via spectral clustering, using the diffusion $K$-means algorithm \cite{LafonLee06}, we quantize the compound space according to geometric structure to create a  common ``codebook'' with centroids $A$, $B$, and $C$. The natural geometry of the data suggests that points near $A$ are closer to points near $B$ than those around $C$. Panel (b), top, shows three histograms, $f$, $g$ and $h$, for distributions over these centroids. The bottom left triangle illustrates how any metric that does not take data geometry into account views the location of the three histograms on a two-dimensional simplex; each of the bins corresponds to a vertex on an equilateral triangle. The arrangement of the bins and the histograms in the bottom right triangle better reflects the interbin relationships. 

Our proposed method takes geometry into account 
by placing the bins (code words) in a Euclidean space defined by a diffusion map. The diffusion map is a rescaled Laplacian eigenmap \cite{BelkinNiyogi03}, 
$\x \in \mathbb{R}^p \mapsto \Psi(\x)  = (\psi_1(\x), \psi_2(\x), \ldots, \psi_d(\x)) \in \mathbb{R}^d$, 
where $\psi_1, \ldots, \psi_d$ are the first $d$ eigenvectors (ordered by their eigenvalues) of a positive semi-definite ``kernel''; in our settings, this kernel could for example be a (renormalized) Gaussian kernel based on cosine distances \cite{Stein1994}
between chemical compounds.
In the diffusion framework, the rescaling is defined such that the Euclidean distance $\| \Psi(\x_i)-\Psi(\x_j) \|$ between any two compounds $i$ and $j$ in the map reflects the intrinsic connectivity structure of the data, defined by 
a Markov random walk (MRW) over the underlying (low-dimensional) data structure.
The details of the construction are given in \cite{LafonLee06}; similar ideas can also be found in \cite{Fouss05, Meila01arandom} 
. The value of having an explicit diffusion metric is that it offers a principled way of linking spectral clustering and manifold learning to data compression and visualization.

{\bf Contribution.} In this work, we extend the diffusion metric between high-dimensional data (chemical compounds) to a generalized diffusion distance (GDD) between weighted sets of such data (smoke samples). Let $\mathbf{f} = \{f_1, \dots, f_K\}$ and $\mathbf{g} = \{g_1, \dots, g_K\}$ be the histogram representations of any two fingerprints, say $S$ and $\widetilde{S}$, where the bin locations are defined by the diffusion $K$-means centers $\{c_i\}_{i=1}^K$  with $K \sim 100$. One can show that the GDD between histograms can be computed very efficiently as  
$GDD(S,\widetilde{S}) = \left\|\sum_{i=1}^K(f_i-g_i) c_i\right\|$,\footnote{The original diffusion distances $D(\x_i,\x_j)$ between compounds $i$ and $j$ correspond to the transition probabilities of a MRW on a graph with the compounds as nodes and the edge weights given by a suitable similarity matrix; the above metric between individual compounds induces a new metric $GDD(S,\widetilde{S})$ between weighted sets $S=\{(\mathbf{x}_1, w_{\mathbf{x}_1}),\dots,(\mathbf{x}_m, w_{\mathbf{x}_m})\}$ and  
	$\widetilde{S}=\{(\widetilde{\mathbf{x}}_1, w_{\widetilde{\mathbf{x}}_1}),\dots,(\widetilde{\mathbf{x}}_m, w_{\widetilde{\mathbf{x}}_{\widetilde{m}}})\}$,
	where generally $m\neq \widetilde{m}$.}
thus with linear complexity $O(K)$.
In comparison, another common set metric such as the Earth Mover Distance (EMD)\cite{Rubner00} has a typical algorithmic complexity of $O(K^3\log K)$ and also does not take the intrinsic geometry of the data distribution into account. Finally, one can show (in prep) that in the limit of $K \rightarrow m$ (or no binning), a version of the GDD converges to the well-known Maximum Mean Discrepancy (MMD) \cite{gretton2012kernel}. The latter result can be used to determine the amount of compression (or the smallest $K$) that would give {\em maximum interpretability, computational efficiency} and {\em low variance} without sacrificing significant accuracy (or bias) relative to MMD.

\subsection*{Aim 2: Differentiating between smoke via kernel machine learning and statistical tests}

GDD 
can serve as input to kernel-based machine learning algorithms  \cite{hofmann2008, scholkopf2018learning} 
for regression and classification of different types of smoke. Our BFRS data include smoke samples collected at air versus ground level. Smoke from the first category escapes the tree line and is more likely to affect communities living near a prescribed burn. Our proposed framework allows us to distinguish between the two different types of smoke, as well as quantify the composition of the smoke most likely to affect regional air quality. In addition to probabilistic classification, we propose a principled method of identifying whether differences in smoke samples are  statistically significant.  For the latter task, we will either employ the local regression test \cite{kim2019} based on a kernel logistic regression \cite{Zhu05} with GDD, or alternatively, test for differences within each of the $K$ partitions via a binomial test \cite{ howell2009statistical, Roeder01}. 

\subsection*{Aim 3: Mapping back to compound space to visualize and explain results}
\begin{figure}
	\floatbox[{\capbeside\thisfloatsetup{capbesideposition={right,top},capbesidewidth=.3\textwidth}}]{figure}[\FBwidth]
	{\caption{\small  Diffusion map of 54 smoke samples from prescribed burns at BFRS.
			Each sample (blue circles for air; green triangles for ground) is represented as a convex combination of
			250 compound ``words'' (orange markers); compare with the histogram vs bin arrangement in Figure 1 right.
			Samples at air versus ground level display differences which are not noticeable
			when using standard techniques to analyze chemical fingerprint data.
		}\label{fig_results_example}}
	{\includegraphics[clip, trim=0.5cm 0cm 1.5cm 1.4cm, width=.47\textwidth]{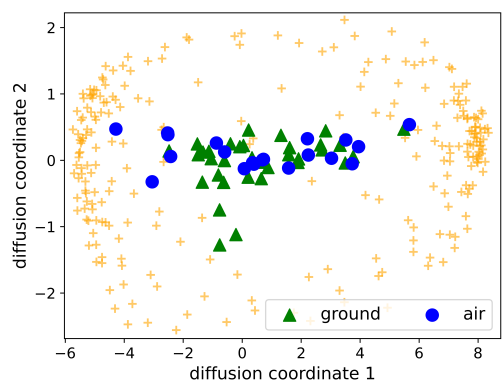}}
\end{figure}

Figure \ref{fig_results_example} illustrates our methodology on our initial BFRS samples.
Once we have defined a common codebook for chemical compounds and represented each chemical fingerprint of smoke as a histogram over the code words (orange markers), we can visualize the smoke samples in diffusion space as points in the convex hull of the $K$ ``words''. This inverse map back to compound space provides insight as to how differences in smoke type (if observed and statistically significant) relate to groups of compounds. After understanding which fuels, burn conditions, and groups of compounds result in certain negative outcomes like poor health or air quality, we will work with forest managers at BFRS and CAL FIRE to develop a forest management plan that reduces climate-induced wildfire risk while minimizing impacts on the regional environment. 


\begin{ack}
	This work is supported in part by the NSF AI Planning Institute for Data-Driven Discovery in Physics, NSF PHY-2020295.
\end{ack}


\begin{thebibliography}{10}
	
	\bibitem{acp-11-4039-2011}
	S.~K. Akagi, R.~J. Yokelson, C.~Wiedinmyer, M.~J. Alvarado, J.~S. Reid,
	T.~Karl, J.~D. Crounse, and P.~O. Wennberg.
	\newblock Emission factors for open and domestic biomass burning for use in
	atmospheric models.
	\newblock {\em Atmospheric Chemistry and Physics}, 11(9):4039--4072, 2011.
	
	\bibitem{BelkinNiyogi03}
	M.~Belkin and P.~Niyogi.
	\newblock Laplacian eigenmaps for dimensionality reduction and data
	representation.
	\newblock {\em Neural Computation}, 15(6):1373--1396, 2003.
	
	\bibitem{Dennison14}
	P.~E. Dennison, S.~C. Brewer, J.~D. Arnold, and M.~A. Moritz.
	\newblock Large wildfire trends in the western {United States}, 1984--2011.
	\newblock {\em Geophysical Research Letters}, 41(8):2928--2933, 2014.
	
	\bibitem{Fouss05}
	F.~Fouss, A.~Pirotte, and M.~Saerens.
	\newblock A novel way of computing similarities between nodes of a graph, with
	application to collaborative recommendation.
	\newblock In {\em Proceedings of the 2005 IEEE/WIC/ACM International Conference
		on Web Intelligence}, WI '05, pages 550--556, 2005.
	
	\bibitem{Goldstein08}
	A.~H. Goldstein, D.~R. Worton, B.~J. Williams, S.~V. Hering, N.~M. Kreisberg,
	O.~Pani{\'c}, and T.~G{\'o}recki.
	\newblock Thermal desorption comprehensive two-dimensional gas chromatography
	for in-situ measurements of organic aerosols.
	\newblock {\em Journal of Chromatography A}, 1186(1-2):340--347, 2008.
	
	\bibitem{gretton2012kernel}
	A.~Gretton, K.~M. Borgwardt, M.~J. Rasch, B.~Sch{\"o}lkopf, and A.~Smola.
	\newblock A kernel two-sample test.
	\newblock {\em The Journal of Machine Learning Research}, 13(1):723--773, 2012.
	
	\bibitem{hofmann2008}
	T.~Hofmann, B.~Sch{\"o}lkopf, and A.~J. Smola.
	\newblock Kernel methods in machine learning.
	\newblock {\em Annals of Statistics}, 36(3):1171--1220, 2008.
	
	\bibitem{howell2009statistical}
	D.~C. Howell.
	\newblock {\em Statistical methods for psychology}.
	\newblock Cengage Learning, 2009.
	
	\bibitem{Jen19}
	C.~N. Jen, L.~E. Hatch, V.~Selimovic, R.~J. Yokelson, R.~Weber, A.~E.
	Fernandez, N.~M. Kreisberg, K.~C. Barsanti, and A.~H. Goldstein.
	\newblock Speciated and total emission factors of particulate organics from
	burning western {US} wildland fuels and their dependence on combustion
	efficiency.
	\newblock {\em Atmospheric Chemistry and Physics}, 19(2):1013--1026, 2019.
	
	\bibitem{Jen18}
	C.~N. Jen, Y.~Liang, L.~E. Hatch, N.~M. Kreisberg, C.~Stamatis, K.~Kristensen,
	J.~J. Battles, S.~L. Stephens, R.~A. York, K.~C. Barsanti, and A.~H.
	Goldstein.
	\newblock High hydroquinone emissions from burning manzanita.
	\newblock {\em Environmental Science \& Technology Letters}, 5(6):309--314,
	2018.
	
	\bibitem{Jolly2015}
	W.~M. Jolly, M.~A. Cochrane, P.~H. Freeborn, Z.~A. Holden, T.~J. Brown, G.~J.
	Williamson, and D.~M. J.~S. Bowman.
	\newblock Climate-induced variations in global wildfire danger from 1979 to
	2013.
	\newblock {\em Nature Communications}, 6(1):7537, 2015.
	
	\bibitem{kim2019}
	I.~Kim, A.~B. Lee, and J.~Lei.
	\newblock Global and local two-sample tests via regression.
	\newblock {\em Electronic Journal of Statistics}, 13(2):5253--5305, 2019.
	
	\bibitem{LafonLee06}
	S.~{Lafon} and A.~B. {Lee}.
	\newblock Diffusion maps and coarse-graining: a unified framework for
	dimensionality reduction, graph partitioning, and data set parameterization.
	\newblock {\em IEEE Transactions on Pattern Analysis and Machine Intelligence},
	28(9):1393--1403, 2006.
	
	\bibitem{Meila01arandom}
	M.~Meila and J.~Shi.
	\newblock A random walks view of spectral segmentation.
	\newblock In {\em Proceedings of AI and Statistics (AISTATS)}, 2001.
	
	\bibitem{Roeder01}
	M.~Roederer and R.~R. Hardy.
	\newblock Frequency difference gating: A multivariate method for identifying
	subsets that differ between samples.
	\newblock {\em Cytometry}, 45(1):56--64, 2001.
	
	\bibitem{Rubner00}
	Y.~Rubner, C.~Tomasi, and L.~J. Guibas.
	\newblock The {Earth Mover's Distance} as a metric for image retrieval.
	\newblock {\em International Journal of Computer Vision}, 40(2):99--121, 2000.
	
	\bibitem{scholkopf2018learning}
	B.~Sch{\"o}lkopf and A.~J. Smola.
	\newblock {\em Learning with kernels: support vector machines, regularization,
		optimization, and beyond}.
	\newblock Adaptive Computation and Machine Learning series, 2018.
	
	\bibitem{Stein1994}
	S.~E. Stein and D.~R. Scott.
	\newblock Optimization and testing of mass spectral library search algorithms
	for compound identification.
	\newblock {\em Journal of the American Society for Mass Spectrometry},
	5(9):859--866, 1994.
	
	\bibitem{Westerling940}
	A.~L. Westerling, H.~G. Hidalgo, D.~R. Cayan, and T.~W. Swetnam.
	\newblock Warming and earlier spring increase western {U.S.} forest wildfire
	activity.
	\newblock {\em Science}, 313(5789):940--943, 2006.
	
	\bibitem{Wiedinmyer11}
	C.~Wiedinmyer, S.~K. Akagi, R.~J. Yokelson, L.~K. Emmons, J.~A. Al--Saadi, J.~J.
	Orlando, and A.~J. Soja.
	\newblock The fire inventory from {NCAR} ({FINN}): a high resolution global
	model to estimate the emissions from open burning.
	\newblock {\em Geoscientific Model Development}, 4(3):625--641, 2011.
	
	\bibitem{Zhu05}
	J.~Zhu and T.~Hastie.
	\newblock Kernel logistic regression and the import vector machine.
	\newblock {\em Journal of Computational and Graphical Statistics},
	14(1):185--205, 2005.
	
\end{thebibliography}
\end{document}